\documentclass[aps,prl,twocolumn,showpacs,a4paper,unsortedaddress,amsmath,amssymb]{revtex4}

\usepackage[latin1]{inputenc} \usepackage{graphicx} \usepackage{dcolumn} \usepackage{bm} \usepackage{hyperref}

\usepackage{subfigure}

\begin{document}

\preprint{ffuov/02-01}

\title{Chiral currents in gold nanotubes}
\author{D. Zs. Manrique$^1$}
\author{J. Cserti$^2$}
\author{C. J. Lambert$^1$}
\thanks{e-mail: c.lambert@lancaster.ac.uk}

\affiliation{$^1$Department of Physics, Lancaster University, Lancaster, LA1 4YB, U. K.}
\affiliation{$^2$Department of Physics of Complex Systems,E{\"o}tv{\"o}s University, H-1117 Budapest, Hungary}


\begin{abstract}
Results are presented for the electron current in gold chiral nanotubes (AuNTs). Starting from the band structure of $(4,3)$ and $(5,3)$ AuNTs, we find that the magnitude of the chiral currents are greater than those found in carbon nanotubes. We also calculate the associated magnetic flux inside the tubes and find this to be higher than the case of carbon nanotubes. Although (4,3) and (5,3) AuNTs carry transverse momenta of similar magnitudes, the low-bias magnetic flux carried by the former is far greater than that carried by the latter. This arises because the low-bias longitudinal current carried by a (4,3) AuNT is significantly smaller than that of a (5,3) AuNT.
\end{abstract}


\maketitle
Nanotubes and nanowires  are  of interest, not only because of their potential for deployment as interconnects, p-n junctions and rectifiers  \cite{1,2}  in future nanoscale circuits, but also because they exhibit fundamental physical properties, such as conductance quantisation\cite{3,4}, and magic numbers reflecting structural stabilities\cite{5,6}.  One ubiquitous property associated with nanotubes is chirality, which arises because there is an infinite number of ways of rolling up a two-dimensional periodic lattice to form a cylinder. In addition to widely-studied  carbon nanotubes  \cite{7},  chiral nanotubes have been formed from a range of other materials, including gold \cite{5, 8,9}, platinum \cite{10}, silver \cite{11}, alkaline metals \cite{12,13,14}  and boron nitride \cite{ 15,16}. These experimental observations have been supported by a range of theoretical investigations \cite{16b,17,18,19,20,21}.

It has recently been noted that the presence of intrinsic chiral electron currents in chiral nanotubes can be exploited to yield photogalvanic effects in heteropolar nanotubes \cite{16b}, a new drive mechanism in carbon-nanotube windmills \cite{22} and to produce internal magnetic fields in carbon nanotube solenoids \cite{23} . In each of these examples, the underlying lattice is hexagonal, with two atoms per unit cell. In contrast nanotubes formed from gold, silver and platinum are derived from triangular lattices with one atom per unit cell and therefore it is of interest examine whether or not these effects are enhanced or diminished compared with their carbon counterparts.

In this paper, to answer these questions, we examine chiral currents in gold nanotubes.  Our choice of gold is in part motivated by the fact that chiral currents are expected to scale with the Fermi velocity of the underlying two-dimensional lattice, which in gold is approximately double that of graphene.

A nanotube formed from a 2D lattice with periodic boundary conditions can be described by a chiral vector  $\mathbf{C} = n \mathbf{a_1} + m \mathbf{a_2}$, which defines the circumference of the nanotube, where $\mathbf{a_1}$,  $\mathbf{a_2}$ are the lattice vectors and $n,m$ are integers. The axis of the nanotube lies parallel to the longitudinal translation vector $\mathbf{T}$ (which is perpendicular the the chiral vector), whose magnitude is equal to the length of the nanotube unit cell, along the tube axis. To understand the currents carried by such a nanotube, we first calculate  the electron group velocity components in the chiral and longitudinal directions. These are given by
$\hbar v_{C} = \partial E(\mathbf{k})/\partial k_{C} $ and $\hbar v_{T} = \partial E(\mathbf{k})/\partial k_{T}$, where $k_C=\mathbf{k}.\mathbf{\hat{C}}$ and $k_T=\mathbf{k}.\mathbf{\hat{T}}$ are components of the wavevector $\mathbf{k}$ parallel to the unit vectors $\mathbf{\hat{C}}$ and $\mathbf{\hat{T}}$ respectively. In the presence of periodic boundary
conditions, $k_C$ is quantized and takes the values
\begin{equation*} k_C^{(q)}=\frac{2\pi q}
{C}, \,\,\,  q=N_0,N_0 + 1, N_0 +2, \ldots, N_0+N-1
\end{equation*} where $N$ is the number of the mini-bands and $C=\vert\mathbf{{C}}\vert$. Since $q$ and $q+N$ are equivalent, $N_0$ is an arbitrary integer. In what follows $N_0$ is chosen such that if $N$ is even, $N_0=1-N/2$, whereas if $N$ is odd, $N_0=\frac{1-N}{2}$. We also choose $N$ to equal to the number of  unit cells of the 2D lattice contained within a unit cell of the nanotube and therefore $-\frac{\pi}{T}\leq k_T<\frac{\pi}{T}$, where $T=\vert\mathbf{{T}}\vert$.


The simplest model describing electronic properties of a triangular lattice consists of a tight-binding Hamiltonian with one orbital per atom. The dispersion relation of such an infinite sheet takes the form
\begin{equation} E(\mathbf{k})=-2\gamma (\cos \mathbf{k}.\mathbf{a_1} + \cos \mathbf{k}.\mathbf{a_2} + \cos \mathbf{k}.\mathbf{a_3}
),
\label{disp1}
\end{equation}
where $\gamma$ is the nearest neighbour coupling, $\mathbf{a_1}=-\frac{a}{2}(1,\sqrt{3})$ and $\mathbf{a_2}=a(1,0)$ are lattice vectors
of the triangular lattice, $\mathbf{a_3}=\mathbf{a_1}+\mathbf{a_2}$ is an auxiliary vector and $a$ is the lattice constant. For a triangular lattice $\mathbf{T} = \frac{2m-n}{d} \mathbf{a_1} + \frac{m-2n}{d}\mathbf{a_2}$, where $d=GCD(2m-n,2n-m)$.
\begin{figure}[ht]
\centering
\includegraphics{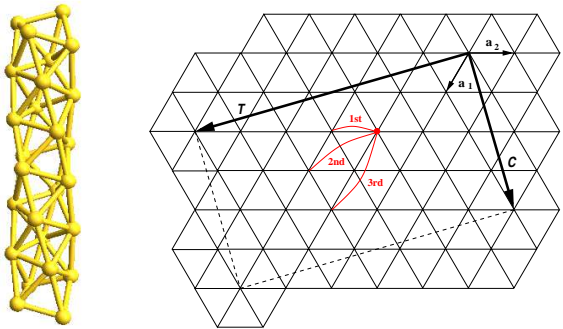}
\caption[]{$(4,3)$ A AuNT and a triangular sheet showing a $(4,3)$ chiral vector. Periodic boundary conditions make the two lattice points at the ends of $\mathbf{C}$ vector identical. The $\mathbf{T}$ vector shows the unit cell of the AuNT. The red lines shows the 1st, 2nd and 3rd nearest neighbours of the highlighted lattice point.}
\label{fig:trigraph}
\end{figure}
For the case of a nanotube formed by imposing periodic boundary conditions, $E(\mathbf{k})=E(k_C^{(q)},k_T)$ and $N=2(n^2+m^2-nm)/d$.
To go beyond this simple model, we performed DFT calculations \cite{siesta,dft} on AuNTs and a triangular gold lattice, whose lattice constant (obtained by relaxing the size of the unit cell) was found to be $a=2.73\text{\AA}$. By fitting the resulting band structures to a third-nearest neighbour model, we obtained a more accurate dispersion relation of the form
\begin{equation}
E(k_C^{(q)},k_T)=\epsilon_0 - 2 \sum_{i}^{9} \gamma_{i} \cos(\alpha_{i} k_C^{(q)} + \beta_{i} k_T)
\label{dispgen}
\end{equation}
where $\alpha_i = \mathbf{\hat{C}}.\mathbf{a_i}$, $\beta_i = \mathbf{\hat{T}}.\mathbf{a_i}$ and $\mathbf{a_4}=\mathbf{a_1}+\mathbf{a_3}$, $\mathbf{a_5}=\mathbf{a_2}+\mathbf{a_3}$, $\mathbf{a_6}=\mathbf{a_2}-\mathbf{a_1}$, $\mathbf{a_7}=2\mathbf{a_1}$, $\mathbf{a_8}=2\mathbf{a_2}$, $\mathbf{a_9}=2\mathbf{a_3}$ auxiliary vectors.

  The values obtained for these couplings are shown in Table \ref{tablegamma}. To account for curvature effects, nine different couplings are needed for AuNTs, whereas, due to symmetry, only three are needed for a flat sheet. With this choice of parameters, the ordering of the bands in eq (2) follows that obtained from DFT. In contrast, we found it impossible to obtain the correct ordering using the simple nearest-neighbour model of eq. (1).

\begin{figure}[ht]
\centering
\subfigure[AuNT $(4,3)$ ]
{
\includegraphics[scale=0.6]{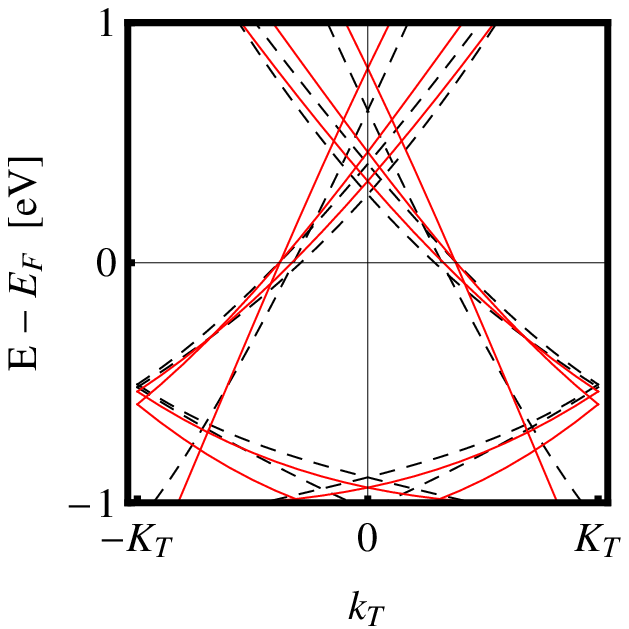}
\label{fig:bands43}
} \subfigure[ AuNT $(5,3)$] {
\includegraphics[scale=0.6]{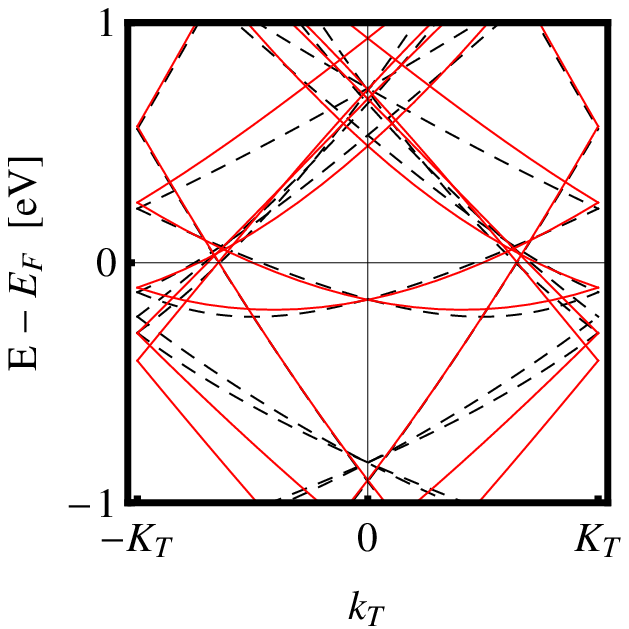}
\label{fig:bands53}}
\caption[]{The black dashed lines show the DFT calculated band structure of the perfect folded AuNT. The red curves show the fitted TB band structure. $E_F$ is the Fermi energy.}
\label{fig:bands}
\end{figure}

In what follows, we focus on the $(5,3)$ and $(4,3)$ AuNTs, since these are realistic experimental targets \cite{6,23b}. As a consequence of curvature, we also expect that the Fermi energy of the tube will be shifted from that of the sheet. This shift is taken into account by an appropriate choice of $\epsilon_0$. Furthermore, since the effect of curvature depends on the choice of $n,m$, the couplings are also allowed to vary with $n,m$. The corresponding band structures are shown in Fig. \ref{fig:bands}.

{\scriptsize }%
\begin{table}
{\tiny }\begin{tabular}{|c|c|c|c|c||c|c|c||c|c|c|}
\hline
{\tiny $\mathbf{n,m}$} & {\tiny $\mathbf{\epsilon_{0}}$} & {\tiny $\mathbf{\gamma_{1}}$} & {\tiny $\mathbf{\gamma_{2}}$} & {\tiny $\mathbf{\gamma_{3}}$} & {\tiny $\mathbf{\gamma_{4}}$} & {\tiny $\mathbf{\gamma_{5}}$} & {\tiny $\mathbf{\gamma_{6}}$} & {\tiny $\mathbf{\gamma_{7}}$} & {\tiny $\mathbf{\gamma_{8}}$} & {\tiny $\mathbf{\gamma_{9}}$}\tabularnewline
\hline
\hline
{\tiny sheet} & {\tiny $\mathbf{0}$} & \multicolumn{3}{c||}{{\tiny $\mathbf{0.95}$}} & \multicolumn{3}{c||}{{\tiny $\mathbf{-0.45}$}} & \multicolumn{3}{c|}{{\tiny $\mathbf{-0.14}$}}\tabularnewline
\hline
{\tiny $\mathbf{4,3}$} & {\tiny $\mathbf{0.6}$} & {\tiny $\mathbf{1.13}$} & {\tiny $\mathbf{0.5}$} & {\tiny $\mathbf{1.56}$} & {\tiny $\mathbf{-0.23}$} & {\tiny $\mathbf{-0.14}$} & {\tiny $\mathbf{-0.4}$} & {\tiny $\mathbf{-0.19}$} & {\tiny $\mathbf{-0.2}$} & {\tiny $\mathbf{-0.13}$}\tabularnewline
\hline
{\tiny $\mathbf{5,3}$} & {\tiny $\mathbf{0.69}$} & {\tiny $\mathbf{0.74}$} & {\tiny $\mathbf{1.07}$} & {\tiny $\mathbf{0.42}$} & {\tiny $\mathbf{-0.61}$} & {\tiny $\mathbf{-0.21}$} & {\tiny $\mathbf{-0.39}$} & {\tiny $\mathbf{0.1}$} & {\tiny $\mathbf{-0.18}$} & {\tiny $\mathbf{0.38}$}\tabularnewline
\hline
\end{tabular}{\tiny \par}
\caption{Fitted coupling parameters for a 2D triangular-lattice sheet with lattice constant $a=2.73\text{\AA}$ and  for perfect $(4,3)$ and $(5,3)$ AuNTs (in eV).}
\label{tablegamma}
\end{table}

By differentiating the dispersion relation (\ref{dispgen}) with respect to $k_C$ and $k_T$, the longitudinal group velocity (parallel to $\mathbf{\hat{T}}$) is found to be
\begin{equation}
 \hbar v_{T}^{(q)} (k_T)= 2 \sum_{i=1}^9 \gamma_i \beta_i \sin(\alpha_i k_C^{(q)}+\beta_i k_T)
\label{labexactvt}
\end{equation}
and in the transverse group velocity (parallel to $\mathbf{\hat{C}}$) is
\begin{equation}
 \hbar v_{C}^{(q)} (k_T)= 2 \sum_{i=1}^9 \gamma_i \alpha_i \sin(\alpha_i k_C^{(q)}+\beta_i k_T)
\label{labexactvc}
\end{equation}

As a reference velocity we note that the Fermi velocity for the triangular sheet(which varies by $\pm 10\%$ around the Fermi surface) has an average value of $v_F=1.8\times 10^6m/s$, when averaged over the Fermi surface.

For an infinitely-long AuNT, we now  compute the chiral velocities of  right-moving electrons, (i.e. with $ v_{T}^{(q)} (k_T) > 0$), by first inverting eq. (\ref{dispgen}) to obtain $k_T$ as a function of $E$ and $q$.  We denote this inverse  $k_T^{(q)+}(E)$, where $+$ sign refers to solutions belonging to branches with $ v_{T}^{(q)} > 0$. Real values of this function arise in the energy range  $\varepsilon_<^{(q)} \leq E< \varepsilon_>^{(q)}$, where $\varepsilon_<^{(q)}$ is the bottom of positive-$ v_{T}^{(q)}$  branch of the $q$th mini-band and $\varepsilon_>^{(q)}$ is the top of the positive-slope branch of the $q$th mini-band. Substituting $k_T^{(q)+}(E)$ into eq. (\ref{labexactvt}) yields the chiral velocity $v_{C}^{(q)+} (E)=v_{C}^{(q)} (k_T^{(q)+}(E))$ belonging to right-moving electrons of energy $E$. The total chiral velocity for the all right-moving electrons of energy $E$ is obtained by summing up the chiral velocities for each mini-band with a real $k_T^{(q)+}(E)$, to yield
\begin{multline}
  v_{C}^{\mathrm{(tot)}+}(E)= \\= \sum_{q=N_0}^{N_0+N-1} v_{C}^{(q)+} (E) \Theta(E-\varepsilon_<^{(q)})\Theta(\varepsilon_>^{(q)}-E)
  \label{vctotal}
\end{multline}


\begin{figure}[ht]
\centering
\includegraphics[scale=0.63]{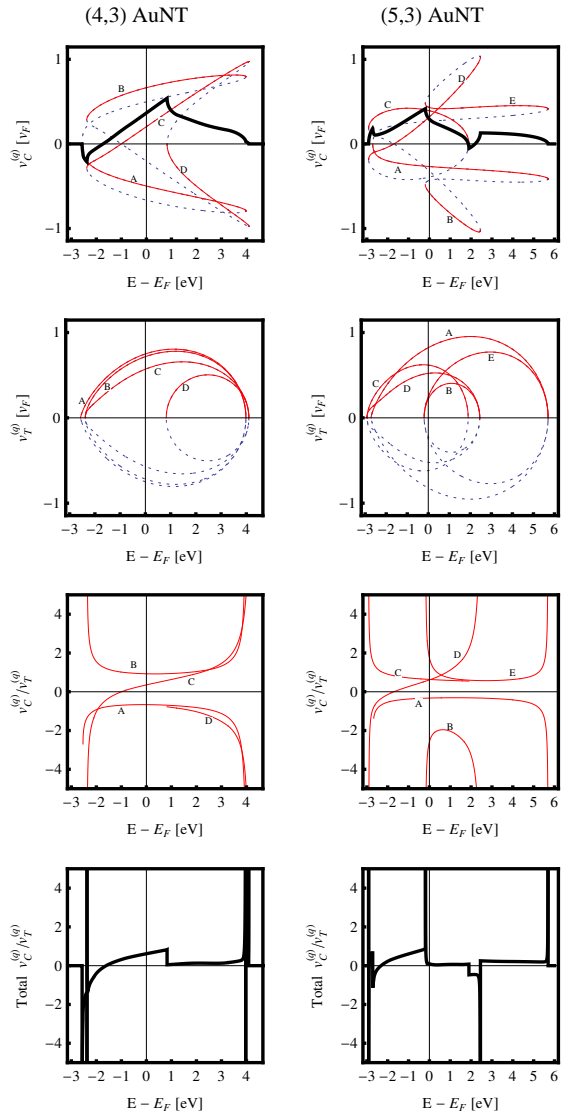}
\caption[]{The top, second and third rows show chiral and longitudinal velocities and their ratios for each miniband of the (4,3) and (5,3) AuNTs. The red curves show those chiral velocity values belonging to right moving electrons. Those of left-moving electrons are shown by dotted curves. The thick black curve shows the sum of the corresponding red curves. and the chiral velocities (red lines) for each channel. The velocities are in units of average Fermi velocity $v_F$ of the triangular lattice sheet. The capital letters label individual open channels.}
\label{fig:evc}
\end{figure}
To invert eq. (\ref{dispgen}), we note that since $\beta_{i} k_T < 2\pi \frac{a}{T}$, and for most of the $n,m$ pairs, $T$ is much longer than $a$, the cosine functions can be approximated by parabolae, from which one can easily obtain an expression for $k_T$ as a function of energy, along with the allowed miniband energy ranges. To obtain an explicit form for the energy dependence of the transverse velocity near an energy band minimum, we note that in the sum on the right-hand-side of eq. (\ref{vctotal}), if the condition ${\varepsilon}_<^{(q)} \leq E< {\varepsilon}_>^{(q)}$ is satisfied for a given $q$, then usually it is not satisfied for $-q$, because a mini-band  with no local extremum in the Brillouin zone satisfies ${\varepsilon}_<^{(\pm q)}={\varepsilon}_>^{(\mp q)}$. {However} if it has local extremum in Brillouin zone, then both $q$ and $-q$ give contributions to the sum because if the miniband has a minimum, then ${\varepsilon}_<^{(q)}={\varepsilon}_<^{(-q)}$, and if it has maximum then ${\varepsilon}_>^{(q)}={\varepsilon}_>^{(-q)}$. Therefore in the case of a minimum, the beginning of the energy ranges of the minibands overlap and in case of maximum the end of the energy ranges overlap. Consequently channels open (close) in pairs when there is a band with local minimum (maximum). Just as a channel pair opens (closes) they give a combined contribution to the right-hand-side of eq. (\ref{vctotal}). Noting that $\beta_i K_T \ll 2\pi$ and Taylor expanding (\ref{labexactvt}) in $k_T$ around zero, one can find that the contribution from the channel pair $q,-q$ is a square-root function of $E$. The $q=0$ case, which belongs to the first channel and does not have partner, yields a contribution
\begin{equation*}
v_{C}^{(0)+}(E)\approx \frac{2}{\hbar}  \sum_i^9 \gamma_i \alpha_i \beta_i  \sqrt{\frac{E-\epsilon_0+2\sum_j^9\gamma_j}{\sum_j^9\gamma_j \beta_j^2}}.
\end{equation*}


\begin{figure}[ht]
\centering
\includegraphics[scale=0.8]{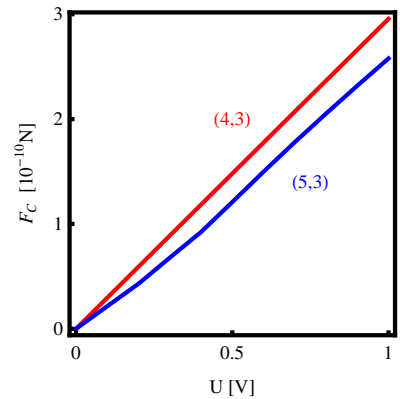}
\caption[]{The forces  against bias voltage. The red curve belongs to $(4,3)$ AuNT and the blue curve belongs to the $(5,3)$ AuNT. }
\label{fig:force}
\end{figure}

\begin{figure}[ht]
\centering
\includegraphics[scale=0.8]{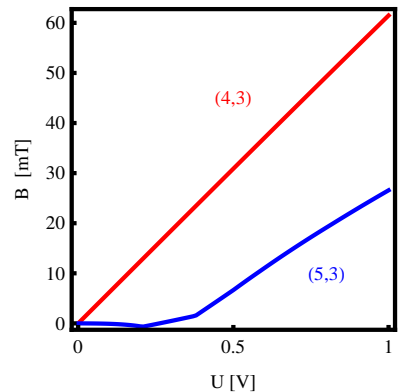}
\caption[]{The magnetic field against bias voltage. The red curves belongs to $(4,3)$ AuNT and the blue curve belongs to the $(5,3)$ AuNT. }
\label{fig:magnetic}
\end{figure}

The above square root behavior near the bottom of a miniband is also found in CNTs \cite{23}. For gold (4,3) and (5,3) AuNTs, the top row of Fig. \ref{fig:evc} shows chiral velocities of individual minibands  (red curves A-D) as function of energy, along with the total chiral velocity (black curves), obtained by summing the individual velocities. The second row of Fig. \ref{fig:evc} shows the longitudinal velocities of individual minibands. The sign of the chiral velocities alternates with successive open channels, leading to oscillations in the total chiral velocity with energy.
To illustrate the behaviour of these quantities when bands open and close, results are plotted over a wider energy range than that used in Fig. \ref{fig:bands}. However results are only meaningful at low energies and therefore when predicting forces and magnetic fields in figures Fig. \ref{fig:force} and Fig. \ref{fig:magnetic}, we revert to energies within $1$ eV of $E_F$.

The chirality of right-moving electrons can play an important role in driving nano motors, because they can exert a torque on the AuNT. An estimate of the maximum possible tangential force is given by the total flux of tangential momentum associated with right-moving electron injected into a bias-voltage window $-\frac{U}{2}\, , \,\frac{U}{2}$. This takes the form
\begin{multline*}
  F_C=
  \frac{2m_e}{h}\int_{-\frac{eU}{2}}^{+\frac{eU}{2}} {v}_{C}^{\mathrm{(tot)}+}(E)d E = \\ = \frac{2m_e}{h}\sum_{q=N_0}^{N_0+N-1} \int_{-\frac{eU}{2}}^{+\frac{eU}{2}} v_{C}^{(q)} (k_T^{(q)+}(E)) \times \\ \times     \Theta(E-\varepsilon_<^{(q)+}) \Theta(\varepsilon_>^{(q)+}-E) d E
\end{multline*}

Fig. \ref{fig:force} shows the resulting force for $(4,3)$ and $(5,3)$ AuNTs. As expected, this is higher than in CNTs, because chiral velocity at the Fermi energy is finite and also, because the Fermi velocity of AuNTs (which sets the velocity scale) is greater than that of CNTs.

A further consequence of the chirality is the presence of an induced magnetic field, given by \cite{23}
\begin{multline*}
  B= \frac{2e}{h} \frac{\mu_0}{C} \sum_{q=N_0}^{N_0+N-1} \int_{-\frac{eU}{2}}^{+\frac{eU}{2}} \frac{v_C^{(q)+}(E)}{v_T^{(q)+}(E)} \times \\ \times \Theta(E-\varepsilon_<^{(q)+}) \Theta(\varepsilon_>^{(q)+}-E) d E .
\end{multline*}
The integrand of this expression involves the ratio of the chiral to longitudinal velocities. For individual minibands, these ratios are shown in the third row of Fig. \ref{fig:evc}. The resulting magnetic fields are shown in Fig. \ref{fig:magnetic}. Comparison between Figs. \ref{fig:force} and \ref{fig:magnetic} shows that although (4,3) and (5,3) AuNTs carry transverse momenta of similar magnitudes, the low-bias magnetic flux carried by a (4,3) AuNT is far greater than that carried by the (5,3) AuNT. This arises because the velocity ratio of the former is significantly higher than the latter, even though they possess similar chiral velocities near the Fermi energy.

We have calculated the chiral velocities carried by electrons in infinitely-long $(4,3)$ and $(5,3)$ AuNTs, which as shown in ref. \cite{23}, can be a guide to the size of chiral currents in finite NTs connected to reservoirs. We have found that in a similar fashion to CNTs, chiral currents are oscillatory functions of energy, but unlike in CNTs, the chiral current has a finite value near $E_F$.  Furthermore the tangential force and the induced magnetic field is higher than in comparable CNTs. We have considered perfectly-periodic nanotubes only. For the future it will be of interest to consider the effect of disorder on chiral currents. In this regard, one notes that at least in one dimension, disorder, which preserves the average spatial symmetry of a lattice, does not completely randomize the phase and density of current-carrying states \cite {cjl1,cjl2}. Therefore, one expects chiral currents to persist in the presence of disorder, although with a diminished magnitude.

Acknowledgements: Supported by the Hungarian Science Foundation OTKA  contracts T48782 \& 75529, EPSRC and the EU ITNs FUNMOLS \& NANOCTM.


\begin{thebibliography}{99}
\bibitem{dft} DFT calculation was carried out within GGA, using a DZP basis set and a force tolerance for coordinate relaxation of 0.01 eV/Ang
\end{thebibliography}
\end{document}